\documentclass[aip, twocolumn,showlabels,showrefs,amsmath,amssymb,pre,superscriptaddress, floatfix, colorst,reprint]{revtex4-1}

\usepackage{mathptmx}

\usepackage[utf8]{inputenc}
\usepackage[T1]{fontenc}
\usepackage{dcolumn}
\usepackage{graphicx}
\usepackage{dcolumn}
\usepackage{bm}
\usepackage{multirow}

\newcommand{\1}{\begin{equation}}
\newcommand{\2}{\end{equation}}
\newcommand{\ea}{\begin{eqnarray}} 
\newcommand{\ee}{\end{eqnarray}}
\newcommand{\4}[2]{{\frac{#1}{#2}}}
 
\newcommand{\Sum}[2]{{\sum\limits_{#1}^{#2}}}

\begin{document}
\title{Which Interactions Dominate in Active Colloids?}

\date{\today}

\author{Benno Liebchen}
\email[]{liebchen@hhu.de}
\affiliation{Institut f\"{u}r Theoretische Physik II: Weiche Materie, Heinrich-Heine-Universit\"{a}t D\"{u}sseldorf, D-40225 D\"{u}sseldorf, Germany}
\author{Hartmut L\"owen}
\affiliation{Institut f\"{u}r Theoretische Physik II: Weiche Materie, Heinrich-Heine-Universit\"{a}t D\"{u}sseldorf, D-40225 D\"{u}sseldorf, Germany}

\begin{abstract}
Despite a mounting evidence that the same gradients which active colloids use for swimming,
induce important cross-interactions (phoretic interaction), 
they are still ignored in most many-body descriptions, perhaps to avoid complexity and a zoo of unknown parameters.
Here we derive a simple model, which reduces phoretic far-field interactions to a pair-interaction 
whose strength is mainly controlled by one genuine parameter (swimming speed).
The model suggests that phoretic interactions 
are generically important for autophoretic colloids (unless effective screening of the phoretic fields is strong) and should 
dominate over hydrodynamic interactions for the typical case of 
half-coating and moderately nonuniform surface mobilities.
Unlike standard minimal models, but in accordance with canonical
experiments, our model generically predicts dynamic clustering in active colloids at low density.
This suggests that dynamic clustering can emerge from the
interplay of screened phoretic attractions and active diffusion.
\end{abstract}


\maketitle

\section{Introduction}
Since their first realization at the turn to the 21st century \cite{Paxton2004,Howse2007},
active colloids \cite{Bechinger2016,Popescu2016,Moran2017}
have evolved from synthetic proof-of-principle microswimmers toward a 
versatile platform for designing functional devices. Now, they are used as
microengines \cite{Kline2005,Golestanian2007,Jiang2010,Bechinger2016,Ebbens2018} and cargo-carriers \cite{Ma2015,Demirors2018}, 
aimed to deliver drugs towards cancer cells in the future, and spark a huge potential for the creation 
of new materials through nonequilibrium self-assembly 
\cite{Palacci2013,Klapp2016,Maggi2016,Zhang2016,Singh2017,Vutukuri2017,Schmidt2018,Aubret2018}. 
These colloids self-propel by catalyzing a chemical reaction on part of their surface,
resulting in a gradient which couples to the surrounding solvent and drives them forward. 
When many active colloids come together, they
self-organize into spectacular patterns, which would be impossible in equilibrium and constitutes their potential for
nonequilibrium self-assembly.  
A typical pattern, reoccurring in canonical experiments with active Janus colloids, 
are so-called living clusters which spontaneously emerge 
at remarkably low densities (area fraction $3-10\%$) and dynamically split up and reform as time proceeds \cite{Theurkauff2012,
Palacci2013,Buttinoni2013,Ginot2018}.  
When trying to understand such collective behaviour in active colloids, 
we are facing complex setups of
motile particles showing
multiple competing interactions, such as steric, hydrodynamic and 
phoretic ones (the latter ones hinge on the cross-action of self-produced chemicals on other colloids). 

Therefore, to reduce complexity and to allow for 
descriptions which are simple enough to 
promote our understanding of the colloids' collective behaviour, 
yet sufficiently realistic to represent typical experimental observations (such as dynamic clustering)
we have to resolve the quest: which interactions dominate in active colloids? 
Presently, the most commonly considered models in the field, like the popular Active Brownian 
particle model \cite{Romanczuk2012,Cates2015} and models involving hydrodynamic interactions \cite{Elgeti2015,Zottl2016}
neglect phoretic interactions altogether, perhaps to avoid complexity and unknown parameters which their description 
usually brings along.
Conversely, recent experiments \cite{Theurkauff2012,Palacci2013,Singh2017,Aubret2018}, simulations \cite{Huang2017,Colberg2017}, and theories \cite{Liebchen2017} 
suggest a crucial importance of phoretic interactions in various active colloids - which, after 15 years of research on active colloids, still 
leaves us with a conflict -- calling for minimal models accounting for phoretic interactions.

Here, our aim is (i) to 
demonstrate that phoretic interactions
are \emph{generically} important in active colloids (unless for strong effective screening) 
and often seem to be the dominant far-field interaction,
(ii) to derive a minimal description of these often neglected interactions, making it easier to account for them in future simulations and theories
and (iii) to show that this minimal description is sufficient to predict 
dynamic clustering, as seen in experiments \cite{Theurkauff2012,Palacci2013,Buttinoni2013,Ginot2018} but not in standard minimal models of active colloids.  
More specifically, we derive the Active Attractive Alignment model (AAA model), 
providing a strongly simplified description of active colloids by 
reducing phoretic interactions to a simple pair interaction among the colloids.
This allows to include them e.g. in Brownian dynamics simulations, rather than requiring 
hydrid particle-field descriptions and releases their modeling from the zoo of unknown parameters it usually involves
\cite{Saha2014, Pohl2014, Meyer2014, Liebchen2015, Liebchen2016,Nejad2018}. 
Remarkably, our derivation shows that the strength of phoretic interactions is 
mainly controlled by one genuine parameter, the self-propulsion speed (or P\'eclet number), rather than involving many unknown parameters. 
This allows to compare the strength of phoretic interactions with hydrodynamic interactions. Our comparison suggest that
phoretic interactions even dominate over 
hydrodynamic interactions for the common case of half-coated Janus colloids with a uniform or a moderately nonuniform surface mobility. 
Thus, as opposed to microswimmers moving by body-shape deformations
\cite{Saintillan2008,Guasto2010,Drescher2010,Elgeti2015,Heidenreich2016,Kaupp2016,Jeanneret2016,Stenhammar2017,Ider2018,Vissers2018},
which are often dominated by hydrodynamic interactions, many active colloids seem to be rather dominated by 
phoretic interactions.
Performing Brownian dynamics simulations we find that 
the AAA model generically predicts dynamic clustering at low density, 
in agreement with experiments \cite{Theurkauff2012,Palacci2013,Buttinoni2013,Ginot2018}, but as opposed to standard minimal models of active colloids. 

Our approach should be broadly useful to model active colloids and to design active self-assembly 
\cite{Soto2014,Gonzalez2018,Singh2017,Aubret2018}. It can be used when the phoretic 
fields relax quasi-instantaneously, which should 
apply to the common case where phoretic interactions are attractive -- in contrast repulsive phoretic interactions can lead to important delay effects 
requiring to explicitly account for the time-evolution of phoretic fields 
\cite{Liebchen2017}.

\section{Phoretic motion in external gradients}
When exposed to a gradient in an imposed phoretic field $c$, 
which may represent e.g. a chemical concentration field, 
the temperature field or an electric potential, colloids move due to 
phoresis. Here, the gradients in $c$ act 
on the fluid elements in the interfacial layer of the colloid and drive a localized solvent flow tangentially to the 
colloidal surface with a velocity, called slip velocity
\1
{\bf v}_s({\bf r}_s)=\mu({\bf r}_s) \nabla_\parallel c({\bf r}_s) \label{slip}
\2 
Here $\mu({\bf r}_s)$ is the phoretic surface mobility,
${\bf r}_s$ points to the colloidal surface (outer edge of interfacial layer) and 
$\nabla_\parallel c$ is the projection of the gradient of $c$ onto the tangential 
plane of the colloid.
The colloid moves opposite to the average surface slip with a velocity 
\cite{Anderson1989}
${\bf v} = \langle -{\bf v}_s({\bf r}_s) \rangle \label{vdrift}$
where brackets represent the average over the colloidal surface. 
If the solvent slips asymmetrically over the colloidal surface, the colloid also rotates with a frequency \cite{Anderson1989}
${\bm \Omega} = \4{3}{2R} \langle {\bf v}_s({\bf r}_s) \times {\bf n}\rangle \label{vrot}$
where $R,{\bf n}$ are the radius and the local surface normal of the colloid. 
Performing surface integrals, and focusing, from now on, on spherical Janus colloids with a catalytic hemisphere with surface mobility 
$\mu_C$, and a mobility of $\mu_N$ on the neutral side, yields: 
\ea
{\bf v}({\bf r})=-\4{\mu_C+\mu_N}{3}\nabla c;\; {\bm \Omega}({\bf r})=\4{3 (\mu_N-\mu_C)}{8R} {\bf e} \times \nabla c \label{Janusresp}
\ee
Here, we have neglected deformations of the field due to the presence of the Janus particle \cite{Bickel2014},
evaluate $c$ at the colloid center ${\bf r}$
for simplicity, and have introduced the unit vector 
${\bf e}$ pointing from the neutral side to the catalytic cap. 

\section{Self-propulsion}
Autophoretic colloidal microswimmers, or active colloids, self-produce phoretic fields 
on part of their surface with a local surface production rate $\sigma({\bf r}_s)$. 
In steady state, we can calculate the field produced by a colloid centered at the origin by solving 
\1
0= D_c \nabla^2 c({\bf r}) + \oint {\rm d}{\bf r}_s \delta\left({\bf r}-{\bf r}_s\right)\sigma({\bf r}_s) -k_d c({\bf r}) \label{chem1}
\2
where the integral is performed over the colloidal surface, $D_c$ is the diffusion coefficient of the relevant phoretic field \cite{footnote2} and 
the sink term $-k_d c$ represents a minimal way to model an effective decay of the phoretic fields, leading to effective screening, 
which may result e.g.
from bulk reactions \cite{Huang2017} (including fuel recovery \cite{Singh2017}) for chemicals and ions. 
While commonly neglected in the literature, Fig.~\ref{fig}G suggests that 
phoretic fields are effectively screened at least for some colloids, which influences phoretic interactions.
(For self-thermophoretic swimmers, $k_d$ might be zero if absorbing boundaries are absent.)
Conversely, self-propulsion, i.e. the phoretic drift of a colloid in its self-produced gradient, depends only on the 
phoretic field close to its surface, so that we can 
ignore the decay. 
Considering a Janus colloid producing chemicals with a local rate
$\sigma=k_0/(2\pi R^2)$ on one hemisphere and $\sigma=0$ on the other one, using
Eqs.~(\ref{slip},\ref{chem1}) for $k_d=0$ and ${\bf v}_0 = \langle -{\bf v}_s({\bf r}_s) \rangle$, we obtain
\cite{Golestanian2007,footnote5} 
\1
{\bf v}_0 = -\4{k_0 (\mu_N+\mu_C)}{16 \pi R^2 D_c} {\bf e} \label{selfprop}
\2
For symmetry reasons the considered Janus colloids do not show self-rotations.

\section{How strong are phoretic interactions?}
Besides leading to self-propulsion, the gradients 
produced by an autophoretic colloid also act in the interfacial layer of all other colloids. 
Here, they drive a solvent
slip over the colloids' surfaces, which induce a phoretic translation and a rotation.  
Following Eqs.~(\ref{slip},\ref{Janusresp},\ref{selfprop}) 
a colloid at the origin causes a 
translation and rotation of a test Janus colloid at position ${\bf r}$ with
\1
{\bf v}_{\rm P}({\bf r}) = -\nu \4{16 \pi R^2 D_c v_0}{3 k_0} \nabla c; \quad {\bm \Omega}_{\rm P}({\bf r}) = \mu_r \4{6\pi D_c R v_0}{k_0} {\bf p} \times \nabla c \label{PI1}
\2
where ${\bf p}$ is the unit vector pointing from ${\bf r}$ into the swimming direction of the test colloid.
Here, 
$\nu=-1$ for swimmers moving with their catalytic cap ahead and $\nu=1$ for cap-behind swimmers \cite{Liebchen2017}; 
we have further used  
$v_0=|{\bf v}_0|$ and have introduced the reduced surface mobility $\mu_r=(\mu_C-\mu_N)/(\mu_C+\mu_R)$.
Now solving Eq.~(\ref{chem1}) in far-field (the integral reduces to $k_0 \delta({\bf r})$), yields the phoretic field produced by the colloid at the origin 
\1
c({\bf r})\approx \4{k_0{\rm e}^{-\kappa r}}{4\pi D_c r}
\label{chemdif}
\2
for $\kappa R/2\ll 1$ and $r\gg R$,
where $\kappa=\sqrt{k_d/D_c}$ is an effective inverse screening length and $\kappa=0$ represents the unscreened case.
(Note that our approach assumes that the phoretic field relaxes quasi-instantaneously to its steady state, 
which is a useful limit for attractive phoretic interactions on which we focus here, 
but can be dangerous at least for the repulsive case \cite{Liebchen2017}.)
Finally combining Eqs.~(\ref{PI1}) and (\ref{chemdif}) yields, in leading order
\ea
{\bf v}_{\rm P}({\bf r})  &=& \4{-4 v_0 R^2 \nu}{3} \nabla \4{{\rm e}^{-\kappa r}}{r}  \label{PIAs} \\ 
{\bm \Omega}_{\rm P}({\bf r}) &=& \4{3v_0 R \mu_r}{2} {\bf p} \times \nabla \4{{\rm e}^{-\kappa r}}{r} \label{PIAs2}
\ee
Except for $\kappa,\mu_r$ which we will estimate below and $\nu=\pm 1$, the prefactors in Eqs.~(\ref{PIAs},\ref{PIAs2}) only depend on the self-propulsion velocity 
and the colloidal radius, which are
well known in experiments. 
We can further see from 
Eq.~(\ref{PIAs}) that colloids at a typical distances of $\sim 5 R$ with $R\sim 1\mu m; v_0\sim 10\mu m/s$, 
approach each other (for $\nu=-1$) within a few seconds due to phoretic interactions (this is consistent with experiments, e.g. 
\cite{Palacci2013,Singh2017}).
For colloids with $R=1\mu m, v_0\sim 10\mu m/s,|\mu_r|=0.15$ \cite{footnote1}, the alignment 
rate with the phoretic gradient produced by an ajacent colloid 
is $|{\bm \Omega}|\sim 0.1/s$, i.e. for the attractive case ($\nu=-1$) colloids may approach each other due to phoretic translation before turning much. 
Thus, it is plausible that when forming dynamic clusters (see below), Janus colloids do not show much orientational order \cite{Ginot2018}.
Still, phoretic alignment should generally play a crucial role for the stability of the uniform phase \cite{Liebchen2017,footnote3}, particularly when 
$|\mu_r| \sim 1$ as e.g. certain thermophoretic swimmers featuring $\mu_C\approx 0$ \cite{Bickel2014}.

\section{Comparison with hydrodynamic interactions}
We now exploit the achieved explicit knowledge of the phoretic interaction coefficients
for a comparison with hydrodynamic interactions. 
\\\textbf{Uniform surface mobility:} 
Besides possible $1/r^2$-contributions which may be led by a small coefficient and are discussed below, 
Janus swimmers always induce a $1/r^3$ flow field, which we now compare with phoretic interactions.
The flow field induced by 
an isotropic (i.e. non-active) colloid in an imposed gradient
at a point ${\bf r}$ relative to its center and well beyond its interfacial layer reads \cite{Morrison1970} ($r:=|{\bf r}|$; $\hat r={\bf r}/r$)
\1
{\bf v}({\bf r}) = \4{1}{2}\left(\4{R}{r}\right)^3 \left(3\hat r \hat r - I \right)\cdot {\bf v}_0 \label{HIphoretic}
\2
The same flow field occurs for Janus colloids with a
uniform surface mobility in a self-produced phoretic gradient, assumed that the colloids cannot distinguish between 
self-produced and imposed phoretic fields. 
Accordingly, this (and similar) flow fields 
commonly occur for Janus colloids (with 
a uniform surface mobility) in the literature \cite{Julicher2009,Zottl2016,Bickel2013,Yang2013,Yang2014,Fedosov2015,Bayati2016,Kreissl2016,Huang2017}.
(Additional flow field contributions may of course arise if the boundary
conditions are different than for a colloid in an imposed gradient \cite{Reigh2016}.) 
We estimate the relative strength of phoretic (\ref{PIAs}) and $1/r^3$-hydrodynamic flows (\ref{HIphoretic}) advecting other colloids (in far field)
via a parameter $m(r) := 8r^3 |\partial_r (\exp{[-\kappa r]}/r)|/(3R)$.
Without a decay of the phoretic field ($\kappa=0$) 
\cite{Palacci2013,Saha2014,Meyer2014,Singh2017}
we have $m \gg 1$ at all relevant distances (i.e. beyond the near field regime) so that phoretic interactions should dominate. 
For $\kappa >0$, hydrodynamic interactions may dominate at very long distances, but rather not at typical ones. 
For $R=1\mu m$ colloids at $10\%$ area fraction (average distance $5.6\mu m$) and $\kappa R=0.25$, we find $m\sim 8.8$, and even for 
$\kappa R\sim 0.5$, we have $m\sim 3.5$); 
higher densities further support phoretic interactions.
Hydrodynamic $1/r^3$-interactions and phoretic interactions 
would break even at distances of $\sim 25R$ for $\kappa R=0.25$ and at $\sim 10R$ for $\kappa R=0.5$.
\\\textbf{Nonuniform surface mobility:} 
Janus swimmers with a non-uniform surface-mobility show additional $1/r^2$ force-dipole contributions 
\cite{Ibrahim2016,Ebbens2018,Popescu2018}, 
whose radial component scales as \cite{Popescu2018}
$v(r) \sim |\mu_r| (R/r)^2 v_0$. 
Thus, for $\kappa=0$, phoretic interactions should be 
$4/(3|\mu_r|)$ times stronger than hydrodynamic $1/r^2$-interactions, at any distance. We roughly 
estimate $1/|\mu_r| \sim 3-20$ for commonly used coating materials \cite{footnote1}, 
so that phoretic interactions seem to dominate. 
Differently, for Janus colloids with a strongly nonuniform surface mobility ($|\mu_r|\sim 1$), which
might apply e.g. to certain electrophoretic swimmers with functionalized surfaces and to
thermophoretic swimmers with thick caps \cite{Bickel2014}
hydrodynamic interactions would be similarly strong as the isotropic component of phoretic interactions. 
If phoretic interactions are screened ($\kappa>0$),
a comparison of $v(r) \sim |\mu_r| (R/r)^2 v_0$ 
with Eq.~(\ref{PIAs}) 
suggests that phoretic and hydrodynamic $1/r^2$-interactions break even at a
distance of $r\approx \left[-1-W(-1,-3|\mu_r|/(4e))\right]/\kappa$ where $W(k,x)$ is the $k$-th branch of the 
Lambert $W$-function (product logarithm). Thus, e.g. for $|\mu_r|=0.2$, phoretic interactions dominate up to a critical distance
of about $3.4/\kappa \approx 13.5R$ for $\kappa R=0.25$, or
at area fractions $>1.7\%$ in uniform suspensions.
\\\textbf{Alignment and Isotropy:} In addition to the pure strength-comparison discussed so far, we note the following: 
(i) Phoretic interactions receive additional support 
from the alignment contribution (at order $\partial_r[\exp(-\kappa r)/r]$), Eq.~(\ref{PIAs2}), which on its own can initiate structure formation 
even at very low density \cite{Liebchen2017}. These alignment contributions are particularly important 
when $|\mu_r|$ is large and might then dominate the collective behaviour of active colloids. 
(ii) Phoretic interactions are isotropic 
(in leading order) and hence superimpose even for randomly oriented particles, whereas anisotropic hydrodynamic flows might mutually cancel to some extend (in bulk).
Possibly, this could additionally support phoretic interactions over hydrodynamic ones and might explain 
why simulations of spherical squirmers involving only hydrodynamic interactions
do not show much structure formation at packing fractions below $\sim 30-40\%$ even for large $|\mu_r|$ \cite{Zottl2014,Blaschke2016}, 
whereas phoretic interactions yield structure formation even at very low density as well will see below. 
These findings are consistent with microscopic simulations of Janus colloids showing clustering at low density due to phoretic interactions, but not 
without \cite{Huang2017}. 
(This does of course not imply, that hydrodynamic interactions essentially average out; 
(rod-shaped) pushers for example are known to destabilize the isotropic phase, at least in the absence of 
rotational diffusion \cite{Saintillan2008,Saintillan2008b}.)
\\\textbf{Limitations:} 
Conversely to the discussed cases, 
hydrodynamic far-field interactions should dominate over phoretic interactions for 
strong effective screening ($\alpha \gg 1$) and in suspensions at very low density ($\lesssim 1-2\%$ or so, depending on $\alpha$ as quantified above).
Hydrodynamic interactions might also be comparatively important for 
significantly nonspherical Janus colloids and for strongly asymmetric coating geometries.
Also in near field, which we do not discuss here, both hydrodynamic and phoretic interactions 
are comparatively involved of course. 
Finally, note that our comparison is based on a simple comparison of pairwise interaction strength, not accounting e.g. for a possible collective impact of momentum conservation due to the solvent; also
our results apply to Janus colloids moving by a self-produced surface slip; in certain swimmers,
e.g. \cite{Buttinoni2013, Lozano2017}, phoretic interactions might be more complicated.

\section{The Active Attractive Aligning Model}
\begin{figure*}
\includegraphics[width=0.9\textwidth]{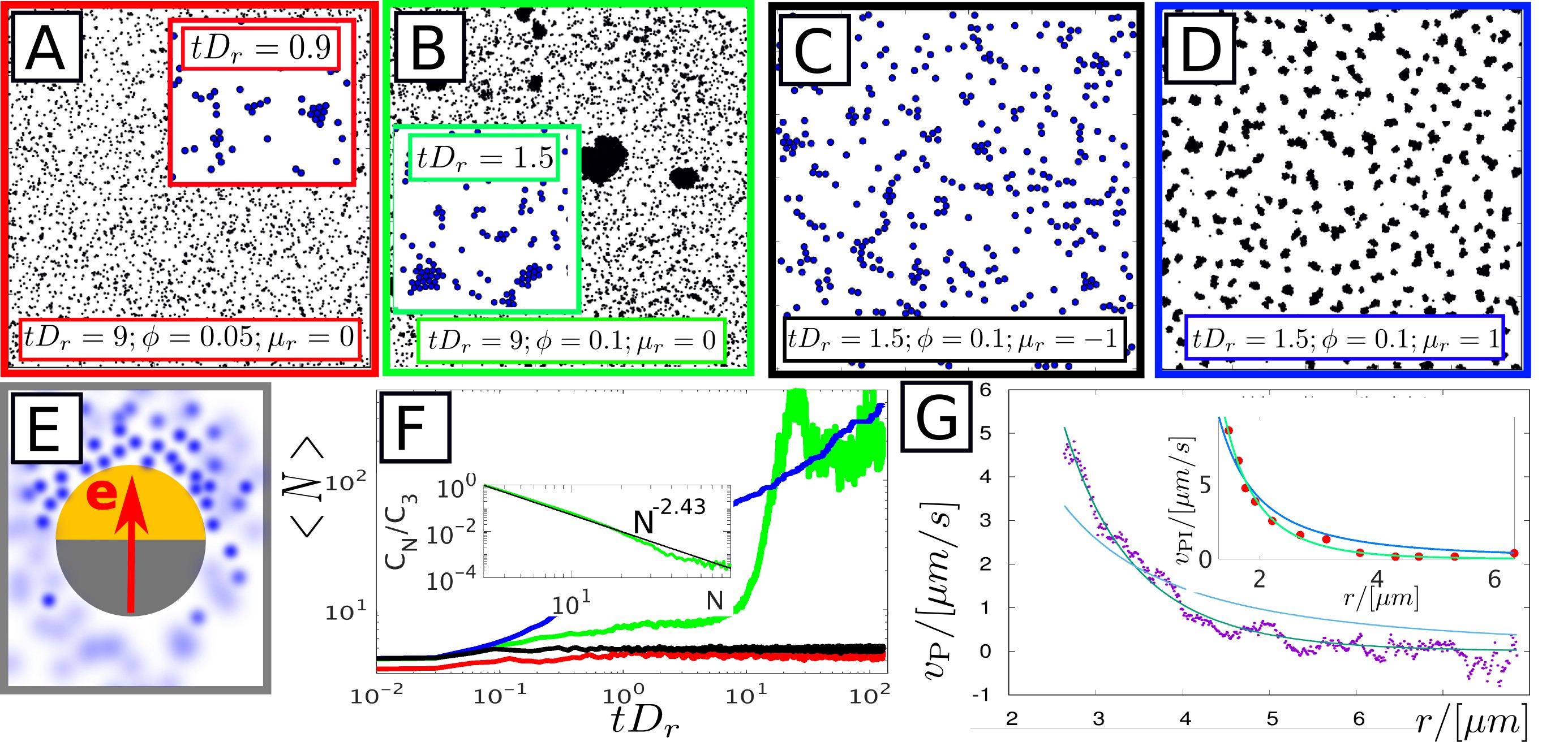}
\caption{\small 
A-D: Dynamic Clustering in the AAA model; snapshots from Brownian dynamics simulations for $N=400-8000$ with 
${\rm Pe}=100, \alpha=0.25, \nu=-1$ at area fractions and times given in the key. 
Panels A-C show dynamic clusters which continuously emerge and split up; 
yielding a finite (nonmacroscopic) cluster size in A,C at late times; 
D shows the system on the way to a 'chemotactic collapse'.
E: Schematic of a Janus colloid swimming with its catalytic cap ahead, hence interacting attractively with other colloids ($\nu=-1$).
F: Time-evolution of the mean cluster size 
calculated by applying a grid with spacing $2x_u$ and counting connected regions;
colors refer to frames in A-D. Inset: Time-averaged cluster size distribution for the data of 
panel B (green) and fit (black) indicating an algebraic decay at small $N$; $C_N/C_3$ is the ratio of $N$-particle clusters  
to 3-particle clusters.
G: Velocity of passive tracers due to the phoretic field produced by Janus colloids in experiments \cite{Singh2017} 
(main figure, dots show our own averages over
tracer trajectories) and \cite{Palacci2013} (inset; dots are based on Fig.~2B in \cite{Palacci2013}). 
Green and blue curves show fits
with and without effective screening respectively. The fits allow for an (upper) 
estimate of $\alpha \lesssim (0.25-0.65)$
in both cases and suggest 
$|\mu|=2\mu_P/(\mu_N+\mu_C) \sim 2-3$ for \cite{Palacci2013}
and $|\mu| \gtrsim 5$ for \cite{Singh2017}, which may however be influenced by additional short-range interactions.}
\label{fig} 
\end{figure*}

To describe the collective behaviour of $N$ active colloids, 
we now consider the Active Brownian particle model as a standard minimal model for active colloids and 
use our previous results to additionally account for phoretic interactions. 
Using $x_u=R$ and $t_u=1/D_r$ as space and time units, where $D_r$ is the translational diffusion constant,
and introducing the P\'eclet number ${\rm Pe}=v_0/(D_r R)$ this 
model reads (in dimensionless units and for colloids moving in quasi-2D):
\1
\dot {\bf x}_i = {\rm Pe}\;{\bf p}_i + {\bf f}_{s}({\bf x}_i); \quad
\dot \theta_i = \sqrt{2}\eta_i(t) \label{ABP}
\2
Eqs.~(\ref{ABP}) describe particles which sterically repel each other (here represented by dimensionless forces ${\bf f}_s$ 
preventing particles to overlap at short distances) 
and self-propel with a velocity $v_0$
in directions 
${\bf p}_i=(\cos\theta_i,\sin\theta_i)$ ($i=1..N$) which change due to rotational Brownian diffusion; 
here $\eta_i$ represents 
Gaussian white noise with zero mean and unit variance.
Following Eq.~(\ref{PIAs},\ref{PIAs2}), 
we can now
account for phoretic far-field interactions 
leading to
the ``Active Attractive Aligning Model'', or AAA model. We define this model for colloids moving in quasi-2D and phoretic fields diffusing in 3D space (see below for a 3D variant and 
\cite{footnote4} for the possible impact of a lower substrate):
\ea
\dot {\bf x}_i &=& {\rm Pe}\; {\bf p}_i - \4{4{\rm Pe}\nu}{3} \nabla u + {\bf f}_{s}({\bf x}_i) \nonumber\\
\dot \theta_i &=& \4{3{\rm Pe} \mu_r}{2}  {\bf p}_i \times \nabla u + \sqrt{2}{\eta}_i(t) \label{AAA}
\ee
Here, $\nabla u=\Sum{j=1; j\neq i}{N}\nabla_{{\bf x}_i}\4{\rm e^{-\alpha x_{ij}}}{x_{ij}}$ with
$x_{ij}=|{\bf x}_i-{\bf x}_j|$ and ${\bf a} \times {\bf b}=a_1 b_2 - a_2 b_1$ for 2D vectors ${\bf a},{\bf b}$
and where we have introduced a screening number 
$\alpha=R\sqrt{k_d/D_c}$.
Remarkably, since we have $\nu=\pm 1$, and expect in many cases $|\mu_r|\ll 1$\cite{footnote1}, for a given screening number 
$\alpha$ (realistic values might be $\alpha \sim 0.25-0.65$, Fig.~\ref{fig}G), the strength of phoretic interactions is mainly controlled 
by one genuine parameter - the P\'eclet number. 
In our simple derivation, we have identified phoretic translations and rotations of the colloids
with formally identical expressions representing reciprocal interaction forces (attractive Yukawa interactions for $\nu=-1$; Coulomb for $\alpha=0$)
and (nonreciprocal) torques aligning the 
self-propulsion direction of the colloids, towards ($\mu_r>0$, positive taxis) or away ($\mu_r<0$, negative taxis) from regions of high particle density. 
The AAA model can be viewed as a description of active colloids containing interactions in leading order in $\mu_r$ (if $|\mu_r|\ll 1$)
\emph{individually} for the center of mass and the orientational dynamics.

\section{Properties of the AAA model}
(i) For $\mu_r= 0,\nu=-1$; the AAA model reduces 
to active Brownian particles with isotropic attractions; however, as opposed to corresponding phenomenological models \cite{Redner2013,Mognetti2013,Prymidis2015,Rein2016,Alarcon2017,Bauerle2018}, 
the AAA model explicitly relates the interaction strength to the P\'eclet number.
Setting $\nu\rightarrow 0$ instead, links the AAA model with the Phoretic Brownian particle model \cite{Liebchen2017} which focuses on phoretic alignment contributions for simplicity, but tracks the 
time-evolution of the phoretic field explicitly.
\cite{Liebchen2017}.
(ii)
The AAA model is based on the assumption that the phoretic fields relax quasi-instantaneously to their steady state.
When they relax slower, which can happen 
even for very large $D_c$ \cite{Liebchen2017}, the phoretic field cannot be eliminated and 
the AAA model becomes invalid; presumably this is relevant mainly for 
repulsive phoretic interactions ($\nu=1$ or $\mu_r<0$) 
\cite{Liebchen2017}.
(iii) 
The Yukawa interactions in Eqs.~(\ref{AAA}) are reciprocal 
only when considering identical colloids.
Mixtures of nonidentical Janus colloids, active-passive mixtures or of 
uniformly coated colloids lead to nonreciprocal interactions
inducing a net motion of pairs \cite{Soto2014,Schmidt2018,Wang2018}.
For example, passive particles can be included in the AAA model 
via 
$\dot {\bf x}^{\rm passive}_i = - (4/3) \mu \nu{\rm Pe} \nabla u$
where ${\rm Pe}$ is the P\'eclet number of the active colloids
and $\mu=2\mu_P/(\mu_N+\mu_C)$ with $\mu_P$ 
being the surface mobility of the (isotropic) passive colloid. 
(iv) For single-specied isotropically coated colloids ($v_0=0$)
the AAA model reduces to the hard-core Yukawa model 
(when accounting for translational diffusion). Thus,
chemically active colloids can be used to realize the (attractive or repulsive) hard-core Yukawa model, 
which has been widely used to describe
effective interactions between charged colloids \cite{Hynninen2003,Heinen2011},
globular proteins \cite{Scholl2013} and fullerenes
\cite{Sun2007}.
(v) Generalizations of the AAA model to 3D are straightforward; here the orientational dynamics follows 
$\dot {\bf p}_i = (3/2) {\rm Pe}\mu_r \left(I-{\bf p}_i {\bf p}_i\right)\nabla u+\sqrt{2}{\bm\eta}_i
\times {\bf p}_i$ where ${\bf p}_i$ is the 3D unit vector representing the swimming direction of particle $i$, ${\bm \eta}_i$ represents Gaussian white noise of zero mean and unit variance and $\times$
now represents the standard 3D cross product.

\section{Dynamic Clustering in the AAA Model}
The AAA model generically leads to dynamic clustering at low density. 
We exemplarily show this 
in Brownian dynamics simulations (Fig.~\ref{fig}), 
at ${\rm Pe}=100$ and $\alpha=0.25$, where we truncate the Yukawa interactions at $16$ particle radii: 
(i) Without alignment ($\mu_r=0$) clusters dynamically emerge, break up and move through space, 
similar as in canonical experiments \cite{Theurkauff2012,Palacci2013,Buttinoni2013,Ginot2018} (see Movie 1).
For an area fraction of $\phi=5\%$, these clusters do not grow beyond a certain size (red line in Fig.~\ref{fig} F). 
Conversely, for $\phi=10\%$ (Movie 2) once a cluster has reached a certain size (Fig.~\ref{fig} B), it continues growing for a comparatively long time
(panel E, green line). However, also here,
the clusters eventually stop growing (at a non-macroscopic size) and dynamically break up leading again to a finite 
average cluster size (Movie 2).
Thus, screened phoretic attractions and active diffusion are sufficient to 
generate dynamic clusters, although phoretic- and other near-field interactions, all neglected here, would of course 
modify the properties of the clusters, once they have emerged.  
(ii) Similarly for $\mu_r=-1$ (strong negative taxis) we also find dynamic clusters (panel C); 
here negative taxis stabilizes the dynamic cluster phase and clusters do not grow at late times 
for $\phi=0.1$ (black curve in F) and also not for $\phi=0.2$ (not shown). 
This combination of attractive translation combined with negative taxis resembles 
\cite{Pohl2014}.
(iii) For $\mu_r=1$ (strong positive taxis) 
we find rigid clusters 
(panel D) which coalesce and form one macrocluster at late times (not shown). 

Note that the clusters seen in cases (i),(ii) differ from those occurring 
as a precursor of motility-induced phase separation \cite{Tailleur2008,Fily2012,Buttinoni2013,Bialke2013,Redner2013,Stenhammar2013,Cates2015,Levis2017}
in the (repulsive) Active Brownian particle (ABP) model 
\cite{Tailleur2008,Fily2012,Buttinoni2013,Bialke2013,Redner2013,Stenhammar2013,Cates2015,Levis2017}. 
The ABP model only leads to very small and short lived clusters at low area fractions;
here the cluster size distribution decays exponentially with the number of particles in the cluster
(unless we are at area fractions of $\gtrsim 30\%$ close to the transition to motility induced phase 
separation). 
In contrast, both in experiments \cite{Ginot2018} and in the AAA model, we see significant clusters at low area fractions ($\leq 10\%$), 
with a cluster size distribution
which decays algebraically at small sizes if the overall area fraction is not too low (inset in panel F). A detailed comparison of cluster sizes and distributions with 
experiments may be performed in future works, 
but might require to account for factors beyond the minimal AAA model, such as 
phoretic and other near-field interactions, a 3D modeling accounting explicitly for a confinement and an understanding of the dependence of $\kappa$ on $v_0$.

\section{Conclusions} 
The derived AAA model provides a minimal description of autophoretic active colloids including phoretic far-field interactions, whose strength we explicitly determine.  
Consequences of our results are as follows: 
(i) The AAA model 
naturally leads to dynamic clustering in the same parameter regime as canonical experiments with active colloids.
This suggests that 
dynamic clustering can occur as a generic result of the interplay of screened phoretic attractions and active diffusion. 
(ii) Phoretic interactions are of crucial importance in typical active colloids. 
In a broad class of autophoretic Janus colloids (half-capped, uniform or moderately nonuniform surface mobility) and corresponding active-passive mixtures, 
they even seem to dominate over the more commonly considered hydrodynamic interactions.
Conversely, hydrodynamic interactions probably dominate over screened phoretic interactions at very low density ($\lesssim 1-2\%$ area fraction, 
depending on $\alpha$) and for cases of strong effective screening ($\alpha \gg 1$).
Finally, for Janus colloids with a strongly asymmetric coating geometry or a strongly nonuniform surface mobility (e.g. thermophoretic swimmers with 
thick caps), phoretic interactions and hydrodynamic interactions 
may be similarly strong. 
Note also that in certain swimmers \cite{Buttinoni2013, Lozano2017}, phoretic interactions might be more complicated than described here.
Future generalizations could account for anisotropy and near-field effects and could explicitly 
account for both hydrodynamic and phoretic interactions to obtain a more general, yet probably more complicated description of active colloids.

\paragraph*{Acknowledgements}
We thank Frederik Hauke for making Fig.~1G (main panel) available and Mihail Popescu and Siegfried Dietrich for 
useful discussions.


\begin{thebibliography}{75}
\expandafter\ifx\csname natexlab\endcsname\relax\def\natexlab#1{#1}\fi
\expandafter\ifx\csname bibnamefont\endcsname\relax
  \def\bibnamefont#1{#1}\fi
\expandafter\ifx\csname bibfnamefont\endcsname\relax
  \def\bibfnamefont#1{#1}\fi
\expandafter\ifx\csname citenamefont\endcsname\relax
  \def\citenamefont#1{#1}\fi
\expandafter\ifx\csname url\endcsname\relax
  \def\url#1{\texttt{#1}}\fi
\expandafter\ifx\csname urlprefix\endcsname\relax\def\urlprefix{URL }\fi
\providecommand{\bibinfo}[2]{#2}
\providecommand{\eprint}[2][]{\url{#2}}


\bibitem[{\citenamefont{Paxton et~al.}(2004)\citenamefont{Paxton, Kistler,
  Olmeda, Sen, St.~Angelo, Cao, Mallouk, Lammert, and Crespi}}]{Paxton2004}
\bibinfo{author}{\bibfnamefont{W.~F.} \bibnamefont{Paxton}},
  \bibinfo{author}{\bibfnamefont{K.~C.} \bibnamefont{Kistler}},
  \bibinfo{author}{\bibfnamefont{C.~C.} \bibnamefont{Olmeda}},
  \bibinfo{author}{\bibfnamefont{A.}~\bibnamefont{Sen}},
  \bibinfo{author}{\bibfnamefont{S.~K.} \bibnamefont{St.~Angelo}},
  \bibinfo{author}{\bibfnamefont{Y.}~\bibnamefont{Cao}},
  \bibinfo{author}{\bibfnamefont{T.~E.} \bibnamefont{Mallouk}},
  \bibinfo{author}{\bibfnamefont{P.~E.} \bibnamefont{Lammert}},
  \bibnamefont{and} \bibinfo{author}{\bibfnamefont{V.~H.}
  \bibnamefont{Crespi}}, \bibinfo{journal}{J. Am. Chem. Soc.}
  \textbf{\bibinfo{volume}{126}}, \bibinfo{pages}{13424}
  (\bibinfo{year}{2004}).

\bibitem[{\citenamefont{Howse et~al.}(2007)\citenamefont{Howse, Jones, Ryan,
  Gough, Vafabakhsh, and Golestanian}}]{Howse2007}
\bibinfo{author}{\bibfnamefont{J.~R.} \bibnamefont{Howse}},
  \bibinfo{author}{\bibfnamefont{R.~A.} \bibnamefont{Jones}},
  \bibinfo{author}{\bibfnamefont{A.~J.} \bibnamefont{Ryan}},
  \bibinfo{author}{\bibfnamefont{T.}~\bibnamefont{Gough}},
  \bibinfo{author}{\bibfnamefont{R.}~\bibnamefont{Vafabakhsh}},
  \bibnamefont{and}
  \bibinfo{author}{\bibfnamefont{R.}~\bibnamefont{Golestanian}},
  \bibinfo{journal}{Phys. Rev. Lett.} \textbf{\bibinfo{volume}{99}},
  \bibinfo{pages}{048102} (\bibinfo{year}{2007}).

\bibitem[{\citenamefont{Bechinger et~al.}(2016)\citenamefont{Bechinger,
  Di~Leonardo, L{\"o}wen, Reichhardt, Volpe, and Volpe}}]{Bechinger2016}
\bibinfo{author}{\bibfnamefont{C.}~\bibnamefont{Bechinger}},
  \bibinfo{author}{\bibfnamefont{R.}~\bibnamefont{Di~Leonardo}},
  \bibinfo{author}{\bibfnamefont{H.}~\bibnamefont{L{\"o}wen}},
  \bibinfo{author}{\bibfnamefont{C.}~\bibnamefont{Reichhardt}},
  \bibinfo{author}{\bibfnamefont{G.}~\bibnamefont{Volpe}}, \bibnamefont{and}
  \bibinfo{author}{\bibfnamefont{G.}~\bibnamefont{Volpe}},
  \bibinfo{journal}{Rev. Mod. Phys.} \textbf{\bibinfo{volume}{88}},
  \bibinfo{pages}{045006} (\bibinfo{year}{2016}).

\bibitem[{\citenamefont{Popescu et~al.}(2016)\citenamefont{Popescu, Uspal, and
  Dietrich}}]{Popescu2016}
\bibinfo{author}{\bibfnamefont{M.~N.} \bibnamefont{Popescu}},
  \bibinfo{author}{\bibfnamefont{W.~E.} \bibnamefont{Uspal}}, \bibnamefont{and}
  \bibinfo{author}{\bibfnamefont{S.}~\bibnamefont{Dietrich}},
  \bibinfo{journal}{Eur. Phys. J. Spec. Top.} \textbf{\bibinfo{volume}{225}},
  \bibinfo{pages}{2189} (\bibinfo{year}{2016}).

\bibitem[{\citenamefont{Moran and Posner}(2017)}]{Moran2017}
\bibinfo{author}{\bibfnamefont{J.~L.} \bibnamefont{Moran}} \bibnamefont{and}
  \bibinfo{author}{\bibfnamefont{J.~D.} \bibnamefont{Posner}},
  \bibinfo{journal}{Ann. Rev. Fluid Mech.} \textbf{\bibinfo{volume}{49}},
  \bibinfo{pages}{511} (\bibinfo{year}{2017}).

\bibitem[{\citenamefont{Kline et~al.}(2005)\citenamefont{Kline, Paxton,
  Mallouk, and Sen}}]{Kline2005}
\bibinfo{author}{\bibfnamefont{T.~R.} \bibnamefont{Kline}},
  \bibinfo{author}{\bibfnamefont{W.~F.} \bibnamefont{Paxton}},
  \bibinfo{author}{\bibfnamefont{T.~E.} \bibnamefont{Mallouk}},
  \bibnamefont{and} \bibinfo{author}{\bibfnamefont{A.}~\bibnamefont{Sen}},
  \bibinfo{journal}{Angew. Chem. Int. Ed.} \textbf{\bibinfo{volume}{44}},
  \bibinfo{pages}{744} (\bibinfo{year}{2005}).

\bibitem[{\citenamefont{Golestanian et~al.}(2007)\citenamefont{Golestanian,
  Liverpool, and Ajdari}}]{Golestanian2007}
\bibinfo{author}{\bibfnamefont{R.}~\bibnamefont{Golestanian}},
  \bibinfo{author}{\bibfnamefont{T.~B.} \bibnamefont{Liverpool}},
  \bibnamefont{and} \bibinfo{author}{\bibfnamefont{A.}~\bibnamefont{Ajdari}},
  \bibinfo{journal}{New J. Phys.} \textbf{\bibinfo{volume}{9}},
  \bibinfo{pages}{126} (\bibinfo{year}{2007}).

\bibitem[{\citenamefont{Jiang et~al.}(2010)\citenamefont{Jiang, Yoshinaga, and
  Sano}}]{Jiang2010}
\bibinfo{author}{\bibfnamefont{H.-R.} \bibnamefont{Jiang}},
  \bibinfo{author}{\bibfnamefont{N.}~\bibnamefont{Yoshinaga}},
  \bibnamefont{and} \bibinfo{author}{\bibfnamefont{M.}~\bibnamefont{Sano}},
  \bibinfo{journal}{Phys. Rev. Lett.} \textbf{\bibinfo{volume}{105}},
  \bibinfo{pages}{268302} (\bibinfo{year}{2010}).

\bibitem[{\citenamefont{Ebbens and Gregory}(2018)}]{Ebbens2018}
\bibinfo{author}{\bibfnamefont{S.~J.} \bibnamefont{Ebbens}} \bibnamefont{and}
  \bibinfo{author}{\bibfnamefont{D.~A.} \bibnamefont{Gregory}},
  \bibinfo{journal}{Acc. Chem. Res.}  (\bibinfo{year}{2018}).

\bibitem[{\citenamefont{Ma et~al.}(2015)\citenamefont{Ma, Hahn, and
  Sanchez}}]{Ma2015}
\bibinfo{author}{\bibfnamefont{X.}~\bibnamefont{Ma}},
  \bibinfo{author}{\bibfnamefont{K.}~\bibnamefont{Hahn}}, \bibnamefont{and}
  \bibinfo{author}{\bibfnamefont{S.}~\bibnamefont{Sanchez}},
  \bibinfo{journal}{J. Am. Chem. Soc.} \textbf{\bibinfo{volume}{137}},
  \bibinfo{pages}{4976} (\bibinfo{year}{2015}).

\bibitem[{\citenamefont{Demir{\"o}rs et~al.}(2018)\citenamefont{Demir{\"o}rs,
  Akan, Poloni, and Studart}}]{Demirors2018}
\bibinfo{author}{\bibfnamefont{A.~F.} \bibnamefont{Demir{\"o}rs}},
  \bibinfo{author}{\bibfnamefont{M.~T.} \bibnamefont{Akan}},
  \bibinfo{author}{\bibfnamefont{E.}~\bibnamefont{Poloni}}, \bibnamefont{and}
  \bibinfo{author}{\bibfnamefont{A.~R.} \bibnamefont{Studart}},
  \bibinfo{journal}{Soft Matter} \textbf{\bibinfo{volume}{14}},
  \bibinfo{pages}{4741} (\bibinfo{year}{2018}).

\bibitem[{\citenamefont{Palacci et~al.}(2013)\citenamefont{Palacci, Sacanna,
  Steinberg, Pine, and Chaikin}}]{Palacci2013}
\bibinfo{author}{\bibfnamefont{J.}~\bibnamefont{Palacci}},
  \bibinfo{author}{\bibfnamefont{S.}~\bibnamefont{Sacanna}},
  \bibinfo{author}{\bibfnamefont{A.~P.} \bibnamefont{Steinberg}},
  \bibinfo{author}{\bibfnamefont{D.~J.} \bibnamefont{Pine}}, \bibnamefont{and}
  \bibinfo{author}{\bibfnamefont{P.~M.} \bibnamefont{Chaikin}},
  \bibinfo{journal}{Science} \textbf{\bibinfo{volume}{339}},
  \bibinfo{pages}{936} (\bibinfo{year}{2013}).

\bibitem[{\citenamefont{Klapp}(2016)}]{Klapp2016}
\bibinfo{author}{\bibfnamefont{S.~H.} \bibnamefont{Klapp}},
  \bibinfo{journal}{Curr. Opin. Colloid Interface Sci.}
  \textbf{\bibinfo{volume}{21}}, \bibinfo{pages}{76} (\bibinfo{year}{2016}).

\bibitem[{\citenamefont{Maggi et~al.}(2016)\citenamefont{Maggi, Simmchen,
  Saglimbeni, Katuri, Dipalo, De~Angelis, Sanchez, and
  Di~Leonardo}}]{Maggi2016}
\bibinfo{author}{\bibfnamefont{C.}~\bibnamefont{Maggi}},
  \bibinfo{author}{\bibfnamefont{J.}~\bibnamefont{Simmchen}},
  \bibinfo{author}{\bibfnamefont{F.}~\bibnamefont{Saglimbeni}},
  \bibinfo{author}{\bibfnamefont{J.}~\bibnamefont{Katuri}},
  \bibinfo{author}{\bibfnamefont{M.}~\bibnamefont{Dipalo}},
  \bibinfo{author}{\bibfnamefont{F.}~\bibnamefont{De~Angelis}},
  \bibinfo{author}{\bibfnamefont{S.}~\bibnamefont{Sanchez}}, \bibnamefont{and}
  \bibinfo{author}{\bibfnamefont{R.}~\bibnamefont{Di~Leonardo}},
  \bibinfo{journal}{Small} \textbf{\bibinfo{volume}{12}}, \bibinfo{pages}{446}
  (\bibinfo{year}{2016}).

\bibitem[{\citenamefont{Zhang et~al.}(2016)\citenamefont{Zhang, Yan, and
  Granick}}]{Zhang2016}
\bibinfo{author}{\bibfnamefont{J.}~\bibnamefont{Zhang}},
  \bibinfo{author}{\bibfnamefont{J.}~\bibnamefont{Yan}}, \bibnamefont{and}
  \bibinfo{author}{\bibfnamefont{S.}~\bibnamefont{Granick}},
  \bibinfo{journal}{Angew. Chem. Int. Ed.} \textbf{\bibinfo{volume}{55}},
  \bibinfo{pages}{5166} (\bibinfo{year}{2016}).

\bibitem[{\citenamefont{Singh et~al.}(2017)\citenamefont{Singh, Choudhury,
  Fischer, and Mark}}]{Singh2017}
\bibinfo{author}{\bibfnamefont{D.~P.} \bibnamefont{Singh}},
  \bibinfo{author}{\bibfnamefont{U.}~\bibnamefont{Choudhury}},
  \bibinfo{author}{\bibfnamefont{P.}~\bibnamefont{Fischer}}, \bibnamefont{and}
  \bibinfo{author}{\bibfnamefont{A.~G.} \bibnamefont{Mark}},
  \bibinfo{journal}{Adv. Mater.} \textbf{\bibinfo{volume}{29}}
  (\bibinfo{year}{2017}).

\bibitem[{\citenamefont{Vutukuri et~al.}(2017)\citenamefont{Vutukuri, Bet,
  Roij, Dijkstra, and Huck}}]{Vutukuri2017}
\bibinfo{author}{\bibfnamefont{H.~R.} \bibnamefont{Vutukuri}},
  \bibinfo{author}{\bibfnamefont{B.}~\bibnamefont{Bet}},
  \bibinfo{author}{\bibfnamefont{R.}~\bibnamefont{Roij}},
  \bibinfo{author}{\bibfnamefont{M.}~\bibnamefont{Dijkstra}}, \bibnamefont{and}
  \bibinfo{author}{\bibfnamefont{W.~T.} \bibnamefont{Huck}},
  \bibinfo{journal}{Sci. Rep.} \textbf{\bibinfo{volume}{7}},
  \bibinfo{pages}{16758} (\bibinfo{year}{2017}).

\bibitem[{\citenamefont{Schmidt et~al.}(2018)\citenamefont{Schmidt, Liebchen,
  L{\"o}wen, and Volpe}}]{Schmidt2018}
\bibinfo{author}{\bibfnamefont{F.}~\bibnamefont{Schmidt}},
  \bibinfo{author}{\bibfnamefont{B.}~\bibnamefont{Liebchen}},
  \bibinfo{author}{\bibfnamefont{H.}~\bibnamefont{L{\"o}wen}},
  \bibnamefont{and} \bibinfo{author}{\bibfnamefont{G.}~\bibnamefont{Volpe}},
  \bibinfo{journal}{arXiv preprint arXiv:1801.06868}  (\bibinfo{year}{2018}).

\bibitem[{\citenamefont{Aubret et~al.}(2018)\citenamefont{Aubret, Youssef,
  Sacanna, and Palacci}}]{Aubret2018}
\bibinfo{author}{\bibfnamefont{A.}~\bibnamefont{Aubret}},
  \bibinfo{author}{\bibfnamefont{M.}~\bibnamefont{Youssef}},
  \bibinfo{author}{\bibfnamefont{S.}~\bibnamefont{Sacanna}}, \bibnamefont{and}
  \bibinfo{author}{\bibfnamefont{J.}~\bibnamefont{Palacci}},
  \bibinfo{journal}{Nat. Phys.}  (\bibinfo{year}{2018}).

\bibitem[{\citenamefont{Huang et~al.}(2017)\citenamefont{Huang, Schofield, and
  Kapral}}]{Huang2017}
\bibinfo{author}{\bibfnamefont{M.-J.} \bibnamefont{Huang}},
  \bibinfo{author}{\bibfnamefont{J.}~\bibnamefont{Schofield}},
  \bibnamefont{and} \bibinfo{author}{\bibfnamefont{R.}~\bibnamefont{Kapral}},
  \bibinfo{journal}{New J. Phys.} \textbf{\bibinfo{volume}{19}},
  \bibinfo{pages}{125003} (\bibinfo{year}{2017}).

\bibitem[{\citenamefont{Colberg et~al.}(2017)\citenamefont{Colberg, and
  Kapral}}]{Colberg2017}
\bibinfo{author}{\bibfnamefont{P.~H.} \bibnamefont{Colberg}},
  \bibnamefont{and} \bibinfo{author}{\bibfnamefont{R.}~\bibnamefont{Kapral}},
  \bibinfo{journal}{J. Chem. Phys.} \textbf{\bibinfo{volume}{147}},
  \bibinfo{pages}{064910} (\bibinfo{year}{2017}).


\bibitem[{\citenamefont{Theurkauff et~al.}(2012)\citenamefont{Theurkauff,
  Cottin-Bizonne, Palacci, Ybert, and Bocquet}}]{Theurkauff2012}
\bibinfo{author}{\bibfnamefont{I.}~\bibnamefont{Theurkauff}},
  \bibinfo{author}{\bibfnamefont{C.}~\bibnamefont{Cottin-Bizonne}},
  \bibinfo{author}{\bibfnamefont{J.}~\bibnamefont{Palacci}},
  \bibinfo{author}{\bibfnamefont{C.}~\bibnamefont{Ybert}}, \bibnamefont{and}
  \bibinfo{author}{\bibfnamefont{L.}~\bibnamefont{Bocquet}},
  \bibinfo{journal}{Phys. Rev. Lett.} \textbf{\bibinfo{volume}{108}},
  \bibinfo{pages}{268303} (\bibinfo{year}{2012}).

\bibitem[{\citenamefont{Buttinoni et~al.}(2013)\citenamefont{Buttinoni,
  Bialk\'{e}, K\"{u}mmel, L\"{o}wen, Bechinger, and Speck}}]{Buttinoni2013}
\bibinfo{author}{\bibfnamefont{I.}~\bibnamefont{Buttinoni}},
  \bibinfo{author}{\bibfnamefont{J.}~\bibnamefont{Bialk\'{e}}},
  \bibinfo{author}{\bibfnamefont{F.}~\bibnamefont{K\"{u}mmel}},
  \bibinfo{author}{\bibfnamefont{H.}~\bibnamefont{L\"{o}wen}},
  \bibinfo{author}{\bibfnamefont{C.}~\bibnamefont{Bechinger}},
  \bibnamefont{and} \bibinfo{author}{\bibfnamefont{T.}~\bibnamefont{Speck}},
  \bibinfo{journal}{Phys. Rev. Lett.} \textbf{\bibinfo{volume}{110}},
  \bibinfo{pages}{238301} (\bibinfo{year}{2013}).

\bibitem[{\citenamefont{Ginot et~al.}(2018)\citenamefont{Ginot, Theurkauff,
  Detcheverry, Ybert, and Cottin-Bizonne}}]{Ginot2018}
\bibinfo{author}{\bibfnamefont{F.}~\bibnamefont{Ginot}},
  \bibinfo{author}{\bibfnamefont{I.}~\bibnamefont{Theurkauff}},
  \bibinfo{author}{\bibfnamefont{F.}~\bibnamefont{Detcheverry}},
  \bibinfo{author}{\bibfnamefont{C.}~\bibnamefont{Ybert}}, \bibnamefont{and}
  \bibinfo{author}{\bibfnamefont{C.}~\bibnamefont{Cottin-Bizonne}},
  \bibinfo{journal}{Nat. Comm.} \textbf{\bibinfo{volume}{9}},
  \bibinfo{pages}{696} (\bibinfo{year}{2018}).

\bibitem[{\citenamefont{Romanczuk et~al.}(2012)\citenamefont{Romanczuk,
  B{\"a}r, Ebeling, Lindner, and Schimansky-Geier}}]{Romanczuk2012}
\bibinfo{author}{\bibfnamefont{P.}~\bibnamefont{Romanczuk}},
  \bibinfo{author}{\bibfnamefont{M.}~\bibnamefont{B{\"a}r}},
  \bibinfo{author}{\bibfnamefont{W.}~\bibnamefont{Ebeling}},
  \bibinfo{author}{\bibfnamefont{B.}~\bibnamefont{Lindner}}, \bibnamefont{and}
  \bibinfo{author}{\bibfnamefont{L.}~\bibnamefont{Schimansky-Geier}},
  \bibinfo{journal}{Eur. Phys. J.} \textbf{\bibinfo{volume}{202}},
  \bibinfo{pages}{1} (\bibinfo{year}{2012}).

\bibitem[{\citenamefont{Cates and Tailleur}(2015)}]{Cates2015}
\bibinfo{author}{\bibfnamefont{M.~E.} \bibnamefont{Cates}} \bibnamefont{and}
  \bibinfo{author}{\bibfnamefont{J.}~\bibnamefont{Tailleur}},
  \bibinfo{journal}{Annu. Rev. Condens. Matter Phys.}
  \textbf{\bibinfo{volume}{6}}, \bibinfo{pages}{219} (\bibinfo{year}{2015}).

\bibitem[{\citenamefont{Elgeti et~al.}(2015)\citenamefont{Elgeti, Winkler, and
  Gompper}}]{Elgeti2015}
\bibinfo{author}{\bibfnamefont{J.}~\bibnamefont{Elgeti}},
  \bibinfo{author}{\bibfnamefont{R.~G.} \bibnamefont{Winkler}},
  \bibnamefont{and} \bibinfo{author}{\bibfnamefont{G.}~\bibnamefont{Gompper}},
  \bibinfo{journal}{Rep. Prog. Phys.} \textbf{\bibinfo{volume}{78}},
  \bibinfo{pages}{056601} (\bibinfo{year}{2015}).

\bibitem[{\citenamefont{Z{\"o}ttl and Stark}(2016)}]{Zottl2016}
\bibinfo{author}{\bibfnamefont{A.}~\bibnamefont{Z{\"o}ttl}} \bibnamefont{and}
  \bibinfo{author}{\bibfnamefont{H.}~\bibnamefont{Stark}}, \bibinfo{journal}{J.
  Phys. Cond. Matter} \textbf{\bibinfo{volume}{28}}, \bibinfo{pages}{253001}
  (\bibinfo{year}{2016}).

\bibitem[{\citenamefont{Liebchen et~al.}(2017)\citenamefont{Liebchen,
  Marenduzzo, and Cates}}]{Liebchen2017}
\bibinfo{author}{\bibfnamefont{B.}~\bibnamefont{Liebchen}},
  \bibinfo{author}{\bibfnamefont{D.}~\bibnamefont{Marenduzzo}},
  \bibnamefont{and} \bibinfo{author}{\bibfnamefont{M.~E.} \bibnamefont{Cates}},
  \bibinfo{journal}{Phys. Rev. Lett.} \textbf{\bibinfo{volume}{118}},
  \bibinfo{pages}{268001} (\bibinfo{year}{2017}).

\bibitem[{\citenamefont{Saintillan and Shelley}(2008)}]{Saintillan2008}
\bibinfo{author}{\bibfnamefont{D.}~\bibnamefont{Saintillan}} \bibnamefont{and}
  \bibinfo{author}{\bibfnamefont{M.~J.} \bibnamefont{Shelley}},
  \bibinfo{journal}{Phys. Rev. Lett.} \textbf{\bibinfo{volume}{100}},
  \bibinfo{pages}{178103} (\bibinfo{year}{2008}).

\bibitem[{\citenamefont{Guasto et~al.}(2010)\citenamefont{Guasto, Johnson, and
  Gollub}}]{Guasto2010}
\bibinfo{author}{\bibfnamefont{J.~S.} \bibnamefont{Guasto}},
  \bibinfo{author}{\bibfnamefont{K.~A.} \bibnamefont{Johnson}},
  \bibnamefont{and} \bibinfo{author}{\bibfnamefont{J.~P.}
  \bibnamefont{Gollub}}, \bibinfo{journal}{Phys. Rev. Lett.}
  \textbf{\bibinfo{volume}{105}}, \bibinfo{pages}{168102}
  (\bibinfo{year}{2010}).

\bibitem[{\citenamefont{Drescher et~al.}(2010)\citenamefont{Drescher,
  Goldstein, Michel, Polin, and Tuval}}]{Drescher2010}
\bibinfo{author}{\bibfnamefont{K.}~\bibnamefont{Drescher}},
  \bibinfo{author}{\bibfnamefont{R.~E.} \bibnamefont{Goldstein}},
  \bibinfo{author}{\bibfnamefont{N.}~\bibnamefont{Michel}},
  \bibinfo{author}{\bibfnamefont{M.}~\bibnamefont{Polin}}, \bibnamefont{and}
  \bibinfo{author}{\bibfnamefont{I.}~\bibnamefont{Tuval}},
  \bibinfo{journal}{Phys. Rev. Lett.} \textbf{\bibinfo{volume}{105}},
  \bibinfo{pages}{168101} (\bibinfo{year}{2010}).

\bibitem[{\citenamefont{Heidenreich et~al.}(2016)\citenamefont{Heidenreich,
  Dunkel, Klapp, and B{\"a}r}}]{Heidenreich2016}
\bibinfo{author}{\bibfnamefont{S.}~\bibnamefont{Heidenreich}},
  \bibinfo{author}{\bibfnamefont{J.}~\bibnamefont{Dunkel}},
  \bibinfo{author}{\bibfnamefont{S.~H.} \bibnamefont{Klapp}}, \bibnamefont{and}
  \bibinfo{author}{\bibfnamefont{M.}~\bibnamefont{B{\"a}r}},
  \bibinfo{journal}{Phys. Rev. E (R)} \textbf{\bibinfo{volume}{94}},
  \bibinfo{pages}{020601} (\bibinfo{year}{2016}).

\bibitem[{\citenamefont{Kaupp and Alvarez}(2016)}]{Kaupp2016}
\bibinfo{author}{\bibfnamefont{U.~B.} \bibnamefont{Kaupp}} \bibnamefont{and}
  \bibinfo{author}{\bibfnamefont{L.}~\bibnamefont{Alvarez}},
  \bibinfo{journal}{Eur. Phys. J. Spec. Top.} \textbf{\bibinfo{volume}{225}},
  \bibinfo{pages}{2119} (\bibinfo{year}{2016}).

\bibitem[{\citenamefont{Jeanneret et~al.}(2016)\citenamefont{Jeanneret,
  Contino, and Polin}}]{Jeanneret2016}
\bibinfo{author}{\bibfnamefont{R.}~\bibnamefont{Jeanneret}},
  \bibinfo{author}{\bibfnamefont{M.}~\bibnamefont{Contino}}, \bibnamefont{and}
  \bibinfo{author}{\bibfnamefont{M.}~\bibnamefont{Polin}},
  \bibinfo{journal}{Eur. Phys. J. Spec. Top.} \textbf{\bibinfo{volume}{225}},
  \bibinfo{pages}{2141} (\bibinfo{year}{2016}).

\bibitem[{\citenamefont{Stenhammar et~al.}(2017)\citenamefont{Stenhammar,
  Nardini, Nash, Marenduzzo, and Morozov}}]{Stenhammar2017}
\bibinfo{author}{\bibfnamefont{J.}~\bibnamefont{Stenhammar}},
  \bibinfo{author}{\bibfnamefont{C.}~\bibnamefont{Nardini}},
  \bibinfo{author}{\bibfnamefont{R.~W.} \bibnamefont{Nash}},
  \bibinfo{author}{\bibfnamefont{D.}~\bibnamefont{Marenduzzo}},
  \bibnamefont{and} \bibinfo{author}{\bibfnamefont{A.}~\bibnamefont{Morozov}},
  \bibinfo{journal}{Phys. Rev. Lett.} \textbf{\bibinfo{volume}{119}},
  \bibinfo{pages}{028005} (\bibinfo{year}{2017}).

\bibitem[{\citenamefont{Daddi-Moussa-Ider
  et~al.}(2018)\citenamefont{Daddi-Moussa-Ider, Lisicki, Mathijssen, Hoell,
  Goh, B{\l}awzdziewicz, Menzel, and L{\"o}wen}}]{Ider2018}
\bibinfo{author}{\bibfnamefont{A.}~\bibnamefont{Daddi-Moussa-Ider}},
  \bibinfo{author}{\bibfnamefont{M.}~\bibnamefont{Lisicki}},
  \bibinfo{author}{\bibfnamefont{A.~J.} \bibnamefont{Mathijssen}},
  \bibinfo{author}{\bibfnamefont{C.}~\bibnamefont{Hoell}},
  \bibinfo{author}{\bibfnamefont{S.}~\bibnamefont{Goh}},
  \bibinfo{author}{\bibfnamefont{J.}~\bibnamefont{B{\l}awzdziewicz}},
  \bibinfo{author}{\bibfnamefont{A.~M.} \bibnamefont{Menzel}},
  \bibnamefont{and}
  \bibinfo{author}{\bibfnamefont{H.}~\bibnamefont{L{\"o}wen}},
  \bibinfo{journal}{J. Phys. Cond. Matter} \textbf{\bibinfo{volume}{30}},
  \bibinfo{pages}{254004} (\bibinfo{year}{2018}).

\bibitem[{\citenamefont{Vissers et~al.}(2018)\citenamefont{Vissers, Brown,
  Koumakis, Dawson, Hermes, Schwarz-Linek, Schofield, French, Koutsos, Arlt
  et~al.}}]{Vissers2018}
\bibinfo{author}{\bibfnamefont{T.}~\bibnamefont{Vissers}},
  \bibinfo{author}{\bibfnamefont{A.~T.} \bibnamefont{Brown}},
  \bibinfo{author}{\bibfnamefont{N.}~\bibnamefont{Koumakis}},
  \bibinfo{author}{\bibfnamefont{A.}~\bibnamefont{Dawson}},
  \bibinfo{author}{\bibfnamefont{M.}~\bibnamefont{Hermes}},
  \bibinfo{author}{\bibfnamefont{J.}~\bibnamefont{Schwarz-Linek}},
  \bibinfo{author}{\bibfnamefont{A.~B.} \bibnamefont{Schofield}},
  \bibinfo{author}{\bibfnamefont{J.~M.} \bibnamefont{French}},
  \bibinfo{author}{\bibfnamefont{V.}~\bibnamefont{Koutsos}},
  \bibinfo{author}{\bibfnamefont{J.}~\bibnamefont{Arlt}}, \bibnamefont{et~al.},
  \bibinfo{journal}{Sci. Adv.} \textbf{\bibinfo{volume}{4}},
  \bibinfo{pages}{eaao1170} (\bibinfo{year}{2018}).

\bibitem[{\citenamefont{Saha et~al.}(2014)\citenamefont{Saha, Golestanian, and
  Ramaswamy}}]{Saha2014}
\bibinfo{author}{\bibfnamefont{S.}~\bibnamefont{Saha}},
  \bibinfo{author}{\bibfnamefont{R.}~\bibnamefont{Golestanian}},
  \bibnamefont{and}
  \bibinfo{author}{\bibfnamefont{S.}~\bibnamefont{Ramaswamy}},
  \bibinfo{journal}{Phys. Rev. E} \textbf{\bibinfo{volume}{89}},
  \bibinfo{pages}{062316} (\bibinfo{year}{2014}).

\bibitem[{\citenamefont{Pohl and Stark}(2014)}]{Pohl2014}
\bibinfo{author}{\bibfnamefont{O.}~\bibnamefont{Pohl}} \bibnamefont{and}
  \bibinfo{author}{\bibfnamefont{H.}~\bibnamefont{Stark}},
  \bibinfo{journal}{Phys. Rev. Lett.} \textbf{\bibinfo{volume}{112}},
  \bibinfo{pages}{238303} (\bibinfo{year}{2014}).

\bibitem[{\citenamefont{Meyer et~al.}(2014)\citenamefont{Meyer,
  Schimansky-Geier, and Romanczuk}}]{Meyer2014}
\bibinfo{author}{\bibfnamefont{M.}~\bibnamefont{Meyer}},
  \bibinfo{author}{\bibfnamefont{L.}~\bibnamefont{Schimansky-Geier}},
  \bibnamefont{and}
  \bibinfo{author}{\bibfnamefont{P.}~\bibnamefont{Romanczuk}},
  \bibinfo{journal}{Phys. Rev. E} \textbf{\bibinfo{volume}{89}},
  \bibinfo{pages}{022711} (\bibinfo{year}{2014}).

\bibitem[{\citenamefont{Liebchen et~al.}(2015)\citenamefont{Liebchen,
  Marenduzzo, Pagonabarraga, and Cates}}]{Liebchen2015}
\bibinfo{author}{\bibfnamefont{B.}~\bibnamefont{Liebchen}},
  \bibinfo{author}{\bibfnamefont{D.}~\bibnamefont{Marenduzzo}},
  \bibinfo{author}{\bibfnamefont{I.}~\bibnamefont{Pagonabarraga}},
  \bibnamefont{and} \bibinfo{author}{\bibfnamefont{M.~E.} \bibnamefont{Cates}},
  \bibinfo{journal}{Phys. Rev. Lett.} \textbf{\bibinfo{volume}{115}},
  \bibinfo{pages}{258301} (\bibinfo{year}{2015}).

\bibitem[{\citenamefont{Liebchen et~al.}(2016)\citenamefont{Liebchen, Cates,
  and Marenduzzo}}]{Liebchen2016}
\bibinfo{author}{\bibfnamefont{B.}~\bibnamefont{Liebchen}},
  \bibinfo{author}{\bibfnamefont{M.~E.} \bibnamefont{Cates}}, \bibnamefont{and}
  \bibinfo{author}{\bibfnamefont{D.}~\bibnamefont{Marenduzzo}},
  \bibinfo{journal}{Soft Matter} \textbf{\bibinfo{volume}{12}},
  \bibinfo{pages}{7259} (\bibinfo{year}{2016}).

\bibitem[{\citenamefont{Nejad et~al.}(2018)\citenamefont{N N}}]{Nejad2018}
\bibinfo{author}{{\bibfnamefont{M.}~\bibnamefont{Nejad}} \bibnamefont{and}
  \bibinfo{author}{\bibfnamefont{A.}~\bibnamefont{Najafi}}},
  \bibinfo{journal}{arXiv preprint arXiv:1712.06004}  (\bibinfo{year}{2018}).


\bibitem[{\citenamefont{Soto and Golestanian}(2014)}]{Soto2014}
\bibinfo{author}{\bibfnamefont{R.}~\bibnamefont{Soto}} \bibnamefont{and}
  \bibinfo{author}{\bibfnamefont{R.}~\bibnamefont{Golestanian}},
  \bibinfo{journal}{Phys. Rev. Lett.} \textbf{\bibinfo{volume}{112}},
  \bibinfo{pages}{068301} (\bibinfo{year}{2014}).

\bibitem[{\citenamefont{Gonzalez and Soto}(2018)}]{Gonzalez2018}
\bibinfo{author}{\bibfnamefont{S.}~\bibnamefont{Gonzalez}} \bibnamefont{and}
  \bibinfo{author}{\bibfnamefont{R.}~\bibnamefont{Soto}}, \bibinfo{journal}{New
  J. Phys.} \textbf{\bibinfo{volume}{20}}, \bibinfo{pages}{053014}
  (\bibinfo{year}{2018}).

\bibitem[{\citenamefont{Anderson}(1989)}]{Anderson1989}
\bibinfo{author}{\bibfnamefont{J.~L.} \bibnamefont{Anderson}},
  \bibinfo{journal}{Ann. Rev. Fluid Mech.} \textbf{\bibinfo{volume}{21}},
  \bibinfo{pages}{61} (\bibinfo{year}{1989}).

  \bibitem[{\citenamefont{Bickel et~al.}(2014)\citenamefont{Bickel, Zecua, and
  W{\"u}rger}}]{Bickel2014}
\bibinfo{author}{\bibfnamefont{T.}~\bibnamefont{Bickel}},
  \bibinfo{author}{\bibfnamefont{G.}~\bibnamefont{Zecua}}, \bibnamefont{and}
  \bibinfo{author}{\bibfnamefont{A.}~\bibnamefont{W{\"u}rger}},
  \bibinfo{journal}{Phys. Rev. E(R)} \textbf{\bibinfo{volume}{89}},
  \bibinfo{pages}{050303} (\bibinfo{year}{2014}).
  
  
\bibitem[{foo({\natexlab{a}})}]{footnote2}
\bibinfo{note}{For self-diffusiophoretic swimmers we understand $c$ as the sum
  of fuel and reaction-product species and $D_c$ as an effective diffusion
  coefficient of the combined field.}

\bibitem[{foo({\natexlab{a}})}]{footnote5}
\bibinfo{note}{ 
Self-thermophoretic swimmers lead to an almost identical expression for the self-propulsion velocity, 
differing only by a constant, e.g. 2/3, depending on the 'field deformation factor'\cite{Bickel2014}.}
  
  
\bibitem[{foo({\natexlab{b}})}]{footnote1}
\bibinfo{note}{
Consider a solute of neutral, dipolar molecules (water, $H_2O_2$)
featuring excluded volume and dipolar interactions with a colloidal surface.
Excluded volume interactions should not depend much on the surface material \cite{Anderson1989} favoring $\mu_r=0$ for Janus particles. 
The surface mobility of a colloid due to dipolar interactions
reads \cite{Anderson1989} $\mu \approx
  \4{-16 kT}{3\eta} \left(\4{\mu_D}{Z e}\right)^2 \xi^2 + \mathcal{O}(\xi^4)$
  where $\xi=\tanh[Z e \zeta/(4 kT)]$ with $\zeta$ being the zeta potential, $\mu_D$ 
the solute dipole moment and $Z$ the valence of the
  support electrolyte (we assume $Z=1$; $Z>1$ reduces
  $|\mu_r|$). Applying these expressions to the halves of a 
Janus colloids with cap (C) and neutral side (N) then suggests
  $|\mu_r|=|(\mu_C-\mu_N)/(\mu_C+\mu_N)|\approx |(\xi_C^2 -
  \xi_N^2)/(\xi_C^2+\xi_N^2)|$. Measurements for typical coating materials:
  $\zeta\approx -64 mV$ both for isotropic $2R=1.7\mu m$ polystyrene (PS)
  and $1\mu m$ gold spheres in $5\% H_2 O_2$ solution \cite{Wang2013}; $\zeta
  \sim -52mV, -67mV$ for polystyrene and silica colloids (sizes $1-3\mu m$)
  respectively in $0.01M$ $KCl$ solution \cite{Ma2015} and for $2R=1\mu m$
  spheres in water $\zeta \sim -50mV$ ($PS,SiO_2$), and $\zeta \sim -63mV$
  ($TiO_2$) \cite{Ni2017}. Measurements for the two halves of a
  $2R=4.8\mu m$ PS-Pt Janus colloid in water (and in $5\%,10\% H_2 O_2$) yield
  $\zeta\sim -95mV (-105mV,-110mV)$ for the $PS$ side and $\zeta\sim -80mV
  (-70mV,-70mV)$ for the $Pt$ \cite{Das2015}. Based on these values we estimate
  $|\mu_r| \sim 0.05-0.3$ for typical Janus swimmers (or less if excluded volume interactions dominate). Similarly, 
for electrophoretic Janus swimmers $|\mu_r|\sim |(\xi_C -
  \xi_N)/(\xi_C+\xi_N)|$ and is therefore significantly smaller than 1 for most typical material combinations.}




\bibitem[{foo({\natexlab{b}})}]{footnote3}
\bibinfo{note}{Instabilities for chemorepulsive colloids at $k_d=0,|\mu_r|=1$ occur 
at about ${\rm Pe}>5$ and area fractions $\sim 10\%$, based on phoretic alignment interactions alone
(see the $k_d=0,|\mu_r|=1$-phase diagram in \cite{Liebchen2017}); 
when e.g. $|\mu_r|=0.2$, they would still occur for typical Janus colloids with ${\rm Pe}>25$, unless effective screening is strong.}



\bibitem[{\citenamefont{Morrison~Jr}(1970)}]{Morrison1970}
\bibinfo{author}{\bibfnamefont{F.}~\bibnamefont{Morrison~Jr}},
  \bibinfo{journal}{J. Colloid Interface Sci.} \textbf{\bibinfo{volume}{34}},
  \bibinfo{pages}{210} (\bibinfo{year}{1970}).




\bibitem[{\citenamefont{J{\"u}licher and Prost}(2009)}]{Julicher2009}
\bibinfo{author}{\bibfnamefont{F.}~\bibnamefont{J{\"u}licher}}
  \bibnamefont{and} \bibinfo{author}{\bibfnamefont{J.}~\bibnamefont{Prost}},
  \bibinfo{journal}{Eur. Phys. J. E} \textbf{\bibinfo{volume}{29}},
  \bibinfo{pages}{27} (\bibinfo{year}{2009}).

\bibitem[{\citenamefont{Bickel et~al.}(2013)\citenamefont{Bickel, Majee, and
  W{\"u}rger}}]{Bickel2013}
\bibinfo{author}{\bibfnamefont{T.}~\bibnamefont{Bickel}},
  \bibinfo{author}{\bibfnamefont{A.}~\bibnamefont{Majee}}, \bibnamefont{and}
  \bibinfo{author}{\bibfnamefont{A.}~\bibnamefont{W{\"u}rger}},
  \bibinfo{journal}{Phys. Rev. E} \textbf{\bibinfo{volume}{88}},
  \bibinfo{pages}{012301} (\bibinfo{year}{2013}).

\bibitem[{\citenamefont{Yang and Ripoll}(2013)}]{Yang2013}
\bibinfo{author}{\bibfnamefont{M.}~\bibnamefont{Yang}} \bibnamefont{and}
  \bibinfo{author}{\bibfnamefont{M.}~\bibnamefont{Ripoll}},
  \bibinfo{journal}{Soft Matter} \textbf{\bibinfo{volume}{9}},
  \bibinfo{pages}{4661} (\bibinfo{year}{2013}).

\bibitem[{\citenamefont{Yang et~al.}(2014)\citenamefont{Yang, Wysocki, and
  Ripoll}}]{Yang2014}
\bibinfo{author}{\bibfnamefont{M.}~\bibnamefont{Yang}},
  \bibinfo{author}{\bibfnamefont{A.}~\bibnamefont{Wysocki}}, \bibnamefont{and}
  \bibinfo{author}{\bibfnamefont{M.}~\bibnamefont{Ripoll}},
  \bibinfo{journal}{Soft Matter} \textbf{\bibinfo{volume}{10}},
  \bibinfo{pages}{6208} (\bibinfo{year}{2014}).

\bibitem[{\citenamefont{Fedosov et~al.}(2015)\citenamefont{Fedosov, Sengupta,
  and Gompper}}]{Fedosov2015}
\bibinfo{author}{\bibfnamefont{D.~A.} \bibnamefont{Fedosov}},
  \bibinfo{author}{\bibfnamefont{A.}~\bibnamefont{Sengupta}}, \bibnamefont{and}
  \bibinfo{author}{\bibfnamefont{G.}~\bibnamefont{Gompper}},
  \bibinfo{journal}{Soft Matter} \textbf{\bibinfo{volume}{11}},
  \bibinfo{pages}{6703} (\bibinfo{year}{2015}).

\bibitem[{\citenamefont{Bayati and Najafi}(2016)}]{Bayati2016}
\bibinfo{author}{\bibfnamefont{P.}~\bibnamefont{Bayati}} \bibnamefont{and}
  \bibinfo{author}{\bibfnamefont{A.}~\bibnamefont{Najafi}},
  \bibinfo{journal}{J. Chem. Phys.} \textbf{\bibinfo{volume}{144}},
  \bibinfo{pages}{134901} (\bibinfo{year}{2016}).

\bibitem[{\citenamefont{Kreissl et~al.}(2016)\citenamefont{Kreissl, Holm, and
  De~Graaf}}]{Kreissl2016}
\bibinfo{author}{\bibfnamefont{P.}~\bibnamefont{Kreissl}},
  \bibinfo{author}{\bibfnamefont{C.}~\bibnamefont{Holm}}, \bibnamefont{and}
  \bibinfo{author}{\bibfnamefont{J.}~\bibnamefont{De~Graaf}},
  \bibinfo{journal}{J. Chem. Phys.} \textbf{\bibinfo{volume}{144}},
  \bibinfo{pages}{204902} (\bibinfo{year}{2016}).

\bibitem[{\citenamefont{Reigh et~al.}(2016)\citenamefont{Re}}]{Reigh2016}
\bibinfo{author}{\bibfnamefont{S.Y.}~\bibnamefont{Reigh}},
\bibinfo{author}{\bibfnamefont{M.-J.}~\bibnamefont{Huang}},
  \bibinfo{author}{\bibfnamefont{J.}~\bibnamefont{Schofield}}, \bibnamefont{and}
  \bibinfo{author}{\bibfnamefont{R.}~\bibnamefont{Kapral}},
  \bibinfo{journal}{Phil. Trans. R. Soc.} \textbf{\bibinfo{volume}{374}},
  \bibinfo{pages}{20160140} (\bibinfo{year}{2016}).



\bibitem[{\citenamefont{Lozano2017}(2017)\citenamefont{Gomez-Solano et al.}}]{Lozano2017}
\bibinfo{author}{\bibfnamefont{J.-R.}~\bibnamefont{Gomez-Solano}},
\bibinfo{author}{\bibfnamefont{S.}~\bibnamefont{Samin}},
  \bibinfo{author}{\bibfnamefont{C.}~\bibnamefont{Lozano}}, 
  \bibinfo{author}{\bibfnamefont{P.}~\bibnamefont{Ruedas-Batuceas}},
  \bibinfo{author}{\bibfnamefont{R.}~\bibnamefont{van Roij}}, \bibnamefont{and}
  \bibinfo{author}{\bibfnamefont{C.}~\bibnamefont{Bechinger}},
  \bibinfo{journal}{Sci. Rep} \textbf{\bibinfo{volume}{7}},
  \bibinfo{pages}{14891} (\bibinfo{year}{2017}).


\bibitem[{\citenamefont{Ibrahim and Liverpool}(2016)}]{Ibrahim2016}
\bibinfo{author}{\bibfnamefont{Y.}~\bibnamefont{Ibrahim}} \bibnamefont{and}
  \bibinfo{author}{\bibfnamefont{T.~B.} \bibnamefont{Liverpool}},
  \bibinfo{journal}{Eur. Phys. J. Spec. Top.} \textbf{\bibinfo{volume}{225}},
  \bibinfo{pages}{1843} (\bibinfo{year}{2016}).

\bibitem[{\citenamefont{Popescu et~al.}(2018)\citenamefont{Popescu, Uspal,
  Eskandari, Tasinkevych, and Dietrich}}]{Popescu2018}
\bibinfo{author}{\bibfnamefont{M.N.}~\bibnamefont{Popescu}},
  \bibinfo{author}{\bibfnamefont{W.E.}~\bibnamefont{Uspal}},
  \bibinfo{author}{\bibfnamefont{Z.}~\bibnamefont{Eskandari}},
  \bibinfo{author}{\bibfnamefont{M.}~\bibnamefont{Tasinkevych}},
  \bibnamefont{and} \bibinfo{author}{\bibfnamefont{S.}~\bibnamefont{Dietrich}},
  \bibinfo{journal}{Eur. Phys. J. E} \textbf{\bibinfo{volume}{41}},
  \bibinfo{pages}{145} (\bibinfo{year}{2018}).

\bibitem[{\citenamefont{Z{\"o}ttl and Stark}(2014)}]{Zottl2014}
\bibinfo{author}{\bibfnamefont{A.}~\bibnamefont{Z{\"o}ttl}} \bibnamefont{and}
  \bibinfo{author}{\bibfnamefont{H.}~\bibnamefont{Stark}},
  \bibinfo{journal}{Phys. Rev. Lett.} \textbf{\bibinfo{volume}{112}},
  \bibinfo{pages}{118101} (\bibinfo{year}{2014}).

\bibitem[{\citenamefont{Blaschke et~al.}(2016)\citenamefont{Blaschke et al.}}]{Blaschke2016}
\bibinfo{author}{\bibfnamefont{J.} \bibnamefont{Blaschke}},
  \bibinfo{author}{\bibfnamefont{M.}~\bibnamefont{Maurer}},
    \bibinfo{author}{\bibfnamefont{K.}~\bibnamefont{Menon}},
  \bibinfo{author}{\bibfnamefont{A.}~\bibnamefont{Z{\"o}ttl}},
  \bibnamefont{and}
  \bibinfo{author}{\bibfnamefont{H.} \bibnamefont{Stark}}, 
  \bibinfo{journal}{Soft Matter} \textbf{\bibinfo{volume}{12}},
  \bibinfo{pages}{9821} (\bibinfo{year}{2016}).
  
\bibitem[{\citenamefont{Saintillan and Shelley}(2008)}]{Saintillan2008b}
\bibinfo{author}{\bibfnamefont{D.}~\bibnamefont{Saintillan}} \bibnamefont{and}
  \bibinfo{author}{\bibfnamefont{M.~J.} \bibnamefont{Shelley}},
  \bibinfo{journal}{Phys. Rev. Fluids} \textbf{\bibinfo{volume}{20}},
  \bibinfo{pages}{123304} (\bibinfo{year}{2008}).

\bibitem[{\citenamefont{Yoshinaga and Liverpool}(2017)}]{Yoshinaga2017}
\bibinfo{author}{\bibfnamefont{N.}~\bibnamefont{Yoshinaga}} \bibnamefont{and}
  \bibinfo{author}{\bibfnamefont{T.~B.} \bibnamefont{Liverpool}},
  \bibinfo{journal}{Phys. Rev. E} \textbf{\bibinfo{volume}{96}},
  \bibinfo{pages}{020603} (\bibinfo{year}{2017}).

  
\bibitem[{foo({\natexlab{a}})}]{footnote4}
\bibinfo{note}{If a lower substrate is present which fully 'reflects' the phoretic field, this can be accounted for by mirror sources, resulting in an additional factor of 2 for the 
strength of phoretic interactions in far-field.}  
  
\bibitem[{\citenamefont{Redner et~al.}(2013)\citenamefont{Redner, Hagan, and
  Baskaran}}]{Redner2013a}
\bibinfo{author}{\bibfnamefont{G.~S.} \bibnamefont{Redner}},
  \bibinfo{author}{\bibfnamefont{A.}~\bibnamefont{Baskaran}},\bibnamefont{and}
  \bibinfo{author}{\bibfnamefont{M.~F.} \bibnamefont{Hagan}}, 
  \bibinfo{journal}{Phys. Rev. E} \textbf{\bibinfo{volume}{88}},
  \bibinfo{pages}{012305} (\bibinfo{year}{2013}).


\bibitem[{\citenamefont{Mognetti et~al.}(2013)\citenamefont{Mognetti}}]{Mognetti2013}
\bibinfo{author}{\bibfnamefont{B.~M.} \bibnamefont{Mognetti}},
  \bibinfo{author}{\bibfnamefont{A.} \bibnamefont{{\v{S}}ari{\'c}}}, \bibnamefont{and}
  \bibinfo{author}{\bibfnamefont{S.}~\bibnamefont{Angioletti-Uberti}},
  \bibinfo{author}{\bibfnamefont{A.}~\bibnamefont{Cacciuto}},
  \bibinfo{author}{\bibfnamefont{C.}~\bibnamefont{Valeriani}},
  \bibinfo{author}{\bibfnamefont{D.}~\bibnamefont{Frenkel}},
  \bibinfo{journal}{Phys. Rev. Lett.} \textbf{\bibinfo{volume}{111}},
  \bibinfo{pages}{245702} (\bibinfo{year}{2013}).


\bibitem[{\citenamefont{Prymidis et~al.}(2015)\citenamefont{Prymidis}}]{Prymidis2015}
\bibinfo{author}{\bibfnamefont{V.}~\bibnamefont{Prymidis}},
  \bibinfo{author}{\bibfnamefont{H.}~\bibnamefont{Sielcken}}, \bibnamefont{and}
  \bibinfo{author}{\bibfnamefont{L.}~\bibnamefont{Filion}},
  \bibinfo{journal}{Soft Matter} \textbf{\bibinfo{volume}{11}},
  \bibinfo{pages}{4158} (\bibinfo{year}{2015}).

\bibitem[{\citenamefont{Rein and Speck}(2016)}]{Rein2016}
\bibinfo{author}{\bibfnamefont{M.}~\bibnamefont{Rein}} \bibnamefont{and}
  \bibinfo{author}{\bibfnamefont{T.}~\bibnamefont{Speck}},
  \bibinfo{journal}{Eur. Phys. J. E} \textbf{\bibinfo{volume}{39}},
  \bibinfo{pages}{84} (\bibinfo{year}{2016}).

\bibitem[{\citenamefont{Alarcon et~al.}(2017)\citenamefont{Alarcon et al}}]{Alarcon2017}
\bibinfo{author}{\bibfnamefont{F.}~\bibnamefont{Alarc{\'o}n}},
  \bibinfo{author}{\bibfnamefont{C.}~\bibnamefont{Valeriani}}, \bibnamefont{and}
  \bibinfo{author}{\bibfnamefont{I.}~\bibnamefont{Pagonabarraga}},
  \bibinfo{journal}{Soft Matter} \textbf{\bibinfo{volume}{13}},
  \bibinfo{pages}{814} (\bibinfo{year}{2017})

\bibitem[{\citenamefont{B\"auerle et~al.}(2018)\citenamefont{B\"auerle,
  Fischer, Speck, and Bechinger}}]{Bauerle2018}
\bibinfo{author}{\bibfnamefont{T.}~\bibnamefont{B\"auerle}},
  \bibinfo{author}{\bibfnamefont{A.}~\bibnamefont{Fischer}},
  \bibinfo{author}{\bibfnamefont{T.}~\bibnamefont{Speck}}, \bibnamefont{and}
  \bibinfo{author}{\bibfnamefont{C.}~\bibnamefont{Bechinger}},
  \bibinfo{journal}{Nat. Comm.} \textbf{\bibinfo{volume}{9}},
  \bibinfo{pages}{3232} (\bibinfo{year}{2018}).

\bibitem[{\citenamefont{Wang et~al.}(2018)\citenamefont{Wang, Popescu, Stavale,
  Ali, Gemming, and Simmchen}}]{Wang2018}
\bibinfo{author}{\bibfnamefont{L.}~\bibnamefont{Wang}},
  \bibinfo{author}{\bibfnamefont{M.~N.} \bibnamefont{Popescu}},
  \bibinfo{author}{\bibfnamefont{F.~L.} \bibnamefont{Stavale}},
  \bibinfo{author}{\bibfnamefont{A.}~\bibnamefont{Ali}},
  \bibinfo{author}{\bibfnamefont{T.}~\bibnamefont{Gemming}}, \bibnamefont{and}
  \bibinfo{author}{\bibfnamefont{J.}~\bibnamefont{Simmchen}},
  \bibinfo{journal}{Soft Matter}  (\bibinfo{year}{2018}).

\bibitem[{\citenamefont{Hynninen and Dijkstra}(2003)}]{Hynninen2003}
\bibinfo{author}{\bibfnamefont{A.-P.} \bibnamefont{Hynninen}} \bibnamefont{and}
  \bibinfo{author}{\bibfnamefont{M.}~\bibnamefont{Dijkstra}},
  \bibinfo{journal}{J. Phys. Cond. Matter} \textbf{\bibinfo{volume}{15}},
  \bibinfo{pages}{S3557} (\bibinfo{year}{2003}).

\bibitem[{\citenamefont{Heinen et~al.}(2011)\citenamefont{Heinen, Banchio, and
  N{\"a}gele}}]{Heinen2011}
\bibinfo{author}{\bibfnamefont{M.}~\bibnamefont{Heinen}},
  \bibinfo{author}{\bibfnamefont{A.~J.} \bibnamefont{Banchio}},
  \bibnamefont{and}
  \bibinfo{author}{\bibfnamefont{G.}~\bibnamefont{N{\"a}gele}},
  \bibinfo{journal}{J. Chem. Phys.} \textbf{\bibinfo{volume}{135}},
  \bibinfo{pages}{154504} (\bibinfo{year}{2011}).

\bibitem[{\citenamefont{Sch{\"o}ll-Paschinger
  et~al.}(2013)\citenamefont{Sch{\"o}ll-Paschinger, Valadez-P{\'e}rez,
  Benavides, and Casta{\~n}eda-Priego}}]{Scholl2013}
\bibinfo{author}{\bibfnamefont{E.}~\bibnamefont{Sch{\"o}ll-Paschinger}},
  \bibinfo{author}{\bibfnamefont{N.~E.} \bibnamefont{Valadez-P{\'e}rez}},
  \bibinfo{author}{\bibfnamefont{A.~L.} \bibnamefont{Benavides}},
  \bibnamefont{and}
  \bibinfo{author}{\bibfnamefont{R.}~\bibnamefont{Casta{\~n}eda-Priego}},
  \bibinfo{journal}{J. Chem. Phys.} \textbf{\bibinfo{volume}{139}},
  \bibinfo{pages}{184902} (\bibinfo{year}{2013}).

\bibitem[{\citenamefont{Sun}(2007)}]{Sun2007}
\bibinfo{author}{\bibfnamefont{J.-X.} \bibnamefont{Sun}},
  \bibinfo{journal}{Phys. Rev. B} \textbf{\bibinfo{volume}{75}},
  \bibinfo{pages}{035424} (\bibinfo{year}{2007}).

\bibitem[{\citenamefont{Tailleur and Cates}(2008)}]{Tailleur2008}
\bibinfo{author}{\bibfnamefont{J.}~\bibnamefont{Tailleur}} \bibnamefont{and}
  \bibinfo{author}{\bibfnamefont{M.}~\bibnamefont{Cates}},
  \bibinfo{journal}{Phys. Rev. Lett.} \textbf{\bibinfo{volume}{100}},
  \bibinfo{pages}{218103} (\bibinfo{year}{2008}).

\bibitem[{\citenamefont{Fily and Marchetti}(2012)}]{Fily2012}
\bibinfo{author}{\bibfnamefont{Y.}~\bibnamefont{Fily}} \bibnamefont{and}
  \bibinfo{author}{\bibfnamefont{M.~C.} \bibnamefont{Marchetti}},
  \bibinfo{journal}{Phys. Rev. Lett.} \textbf{\bibinfo{volume}{108}},
  \bibinfo{pages}{235702} (\bibinfo{year}{2012}).

\bibitem[{\citenamefont{Bialk{\'e} et~al.}(2013)\citenamefont{Bialk{\'e},
  L{\"o}wen, and Speck}}]{Bialke2013}
\bibinfo{author}{\bibfnamefont{J.}~\bibnamefont{Bialk{\'e}}},
  \bibinfo{author}{\bibfnamefont{H.}~\bibnamefont{L{\"o}wen}},
  \bibnamefont{and} \bibinfo{author}{\bibfnamefont{T.}~\bibnamefont{Speck}},
  \bibinfo{journal}{Europhys. Lett.} \textbf{\bibinfo{volume}{103}},
  \bibinfo{pages}{30008} (\bibinfo{year}{2013}).

\bibitem[{\citenamefont{Redner et~al.}(2013)\citenamefont{Redner, Hagan, and
  Baskaran}}]{Redner2013}
\bibinfo{author}{\bibfnamefont{G.~S.} \bibnamefont{Redner}},
  \bibinfo{author}{\bibfnamefont{M.~F.} \bibnamefont{Hagan}}, \bibnamefont{and}
  \bibinfo{author}{\bibfnamefont{A.}~\bibnamefont{Baskaran}},
  \bibinfo{journal}{Phys. Rev. Lett.} \textbf{\bibinfo{volume}{110}},
  \bibinfo{pages}{055701} (\bibinfo{year}{2013}).



\bibitem[{\citenamefont{Stenhammar et~al.}(2013)\citenamefont{Stenhammar,
  Tiribocchi, Allen, Marenduzzo, and Cates}}]{Stenhammar2013}
\bibinfo{author}{\bibfnamefont{J.}~\bibnamefont{Stenhammar}},
  \bibinfo{author}{\bibfnamefont{A.}~\bibnamefont{Tiribocchi}},
  \bibinfo{author}{\bibfnamefont{R.~J.} \bibnamefont{Allen}},
  \bibinfo{author}{\bibfnamefont{D.}~\bibnamefont{Marenduzzo}},
  \bibnamefont{and} \bibinfo{author}{\bibfnamefont{M.~E.} \bibnamefont{Cates}},
  \bibinfo{journal}{Phys. Rev. Lett.} \textbf{\bibinfo{volume}{111}},
  \bibinfo{pages}{145702} (\bibinfo{year}{2013}).

\bibitem[{\citenamefont{Levis et~al.}(2017)\citenamefont{Levis, Codina, and
  Pagonabarraga}}]{Levis2017}
\bibinfo{author}{\bibfnamefont{D.}~\bibnamefont{Levis}},
  \bibinfo{author}{\bibfnamefont{J.}~\bibnamefont{Codina}}, \bibnamefont{and}
  \bibinfo{author}{\bibfnamefont{I.}~\bibnamefont{Pagonabarraga}},
  \bibinfo{journal}{Soft Matter} \textbf{\bibinfo{volume}{13}},
  \bibinfo{pages}{8113} (\bibinfo{year}{2017}).

\bibitem[{\citenamefont{Wang et~al.}(2013)\citenamefont{Wang, Duan, Sen, and
  Mallouk}}]{Wang2013}
\bibinfo{author}{\bibfnamefont{W.}~\bibnamefont{Wang}},
  \bibinfo{author}{\bibfnamefont{W.}~\bibnamefont{Duan}},
  \bibinfo{author}{\bibfnamefont{A.}~\bibnamefont{Sen}}, \bibnamefont{and}
  \bibinfo{author}{\bibfnamefont{T.~E.} \bibnamefont{Mallouk}},
  \bibinfo{journal}{Proc. Natl. Acad. Sci.} p. \bibinfo{pages}{201311543}
  (\bibinfo{year}{2013}).

\bibitem[{\citenamefont{Ni et~al.}(2017)\citenamefont{Ni, Marini, Buttinoni,
  Wolf, and Isa}}]{Ni2017}
\bibinfo{author}{\bibfnamefont{S.}~\bibnamefont{Ni}},
  \bibinfo{author}{\bibfnamefont{E.}~\bibnamefont{Marini}},
  \bibinfo{author}{\bibfnamefont{I.}~\bibnamefont{Buttinoni}},
  \bibinfo{author}{\bibfnamefont{H.}~\bibnamefont{Wolf}}, \bibnamefont{and}
  \bibinfo{author}{\bibfnamefont{L.}~\bibnamefont{Isa}}, \bibinfo{journal}{Soft
  Matter} \textbf{\bibinfo{volume}{13}}, \bibinfo{pages}{4252}
  (\bibinfo{year}{2017}).

\bibitem[{\citenamefont{Das et~al.}(2015)\citenamefont{Das}}]{Das2015}
\bibinfo{author}{\bibfnamefont{S.}~\bibnamefont{Das}},
  \bibinfo{author}{\bibfnamefont{G.}~\bibnamefont{Astha}},
  \bibinfo{author}{\bibfnamefont{A.I.}~\bibnamefont{Campbell}},
  \bibinfo{author}{\bibfnamefont{J.}~\bibnamefont{Howse}},
  \bibinfo{author}{\bibfnamefont{A.}~\bibnamefont{Sen}},
  \bibinfo{author}{\bibfnamefont{D.}~\bibnamefont{Velegol}},
  \bibinfo{author}{\bibfnamefont{R.}~\bibnamefont{Golestanian}}, \bibnamefont{and}
  \bibinfo{author}{\bibfnamefont{S.J.}~\bibnamefont{Ebbens}}, \bibinfo{journal}{Nat. Comm.} \textbf{\bibinfo{volume}{6}}, \bibinfo{pages}{8999}
  (\bibinfo{year}{2015}).


%



\end{thebibliography}

\end{document}